\documentclass[11pt]{article}
\usepackage{cospar}
\usepackage{natbib}
\usepackage{url}

\usepackage{graphicx}
\usepackage[figuresright]{rotating}

\newcommand{\xte}{\textit{RXTE}}

\newcommand{\tfe}{1E~1048.1--5937}
\newcommand{\tfn}{1E~2259.1+586}

\newcommand{\oft}{4U~0142+61}

\title{Anomalous X-ray Pulsars: \\ \textit{Long-Term Monitoring} \\ and \\ \textit{Soft-Gamma Repeater like X-ray Bursts}}

\author{F.~P.~Gavriil \address{Physics Department, McGill University, Montreal, QC H3A 2T8, Canada},
        V.~M.~Kaspi$^{1}$,
        and
        P.~M.~Woods\address{Space Science Research Center, National Space Science
and Technology Center, Huntsville, AL 35805, USA; Universities Space Research Association}}

\begin{document}
% typeset front matter   
\maketitle

\begin{abstract}
We report on long-term monitoring of anomalous X-ray pulsars (AXPs)
using the \textit{Rossi X-ray Timing Explorer} (\xte). 
Using phase-coherent timing, we find a wide variety of 
behaviors among the sources, ranging
from high stability (in \tfn\ in quiescence and \oft), to instabilities so severe
that phase-coherent timing is not possible (in \tfe).
 We note a
correlation in which timing stability in AXPs decreases with increasing
$\dot{\nu}$.  The timing stability of soft gamma repeaters (SGRs) in
quiescence is consistent with this trend, which is similar to one seen in 
radio pulsars. We find no
significant pulse morphology variations in any AXP in quiescence.  
We considered  high signal-to-noise average pulse profiles for
each AXP as a function of energy.  We show that, as
in the timing properties, there is a variety of different behaviors for
the energy dependence.  
We also used the monitoring and archival data to obtain pulsed flux
time series for each source.  We have found no large changes in pulsed
flux for any source in quiescence,  and have set $1\sigma$ upper limits on variations
$\sim$20--30\% depending on the source.
We have recently discovered  bursts from the direction of two  AXPs: \tfe\ the most SGR-like AXP,  and \tfn\ the most rotationally stable AXP. We compare the temporal, 
spectral and flux properties of these events  to those of SGR bursts, and show  that 
the two phenomena are very similar. These results imply
a close relationship between AXPs and SGRs, with both being magnetars.
\end{abstract}

\section*{INTRODUCTION}

The nature of anomalous X-ray pulsars (AXPs) was a mystery for the past
20 years, since the discovery of the first example \cite[\tfn;][]{fg81}.
There are currently only five confirmed AXPs, all of which are located in
the Galactic plane, with two at the geometric centers of
supernova remnants \citep{fg81,vg97}.  AXP characteristics can be summarized as
follows \cite[see][for a review]{ims02}:
they exhibit X-ray
pulsations in the range $L_x \sim$6--12~s; they have pulsed X-ray
luminosities in the range $\sim 10^{33}-10^{35}$~erg~s$^{-1}$; they
spin down regularly;  they have spectra that are characterized by
thermal emission of $kT \sim 0.4$~keV plus a hard power-law tail. They have 
faint or no optical/IR counterparts \citep{hkvk00,hvk00,ims02}. The optical counterpart of one AXP, \oft\, is reportedly pulsed \citep{km02}.

AXPs are called ``anomalous'' because it has been  unclear what powers their
radiation.  They are not rotation-powered as their
observed X-ray luminosities are much greater than the rate of loss of
rotational kinetic energy inferred from their spin-down. 
They were long thought to be accreting from a low
mass companion \cite[e.g.][]{ms95}.  However this model is
difficult to reconcile with observations: the absence of Doppler
shifts even on short time scales \cite[e.g.][]{mis98}, the absence
of a detectable optical/IR companion or bright accretion disk \cite[see][]{ms95,hkvk00},
the apparent associations with supernova remnants, 
that AXP spectra are very different from those of
known accreting sources, and that $L_x$ is generally smaller than in
known accreting sources, all are inconsistent with this scenario.
\citet{chn00} \cite[see also][]{vtv95,cso+95}
proposed that AXPs are accreting from a fossil disk made of
fall-back material from the supernova.
However, this model overpredicts the observed optical/IR
flux from the disk \citep{phn00,hkvk00,hvk00,kkk+01}.

\citet{td96a} suggested that the main source of free energy for AXP
emission is from the magnetic field itself. In
this model, AXPs are young, isolated, highly magnetized neutron stars
or ``magnetars''.  Prior to our {\it RXTE} monitoring
campaign, the primary evidence in favor of this model for AXPs was the
inferred strength of the surface dipolar magnetic field required to
slow the pulsar down {\it in vacuo}, assuming magnetic dipole braking
as in radio pulsars.  These fields are in the range $\sim
10^{14}-10^{15}$~G.  The spin-down ages in this model, inferred
assuming a small birth spin period, are in the range
$\sim$8--200~kyr.  This suggested youth is further evidenced by the 
associations with supernova remnants, and
from the location of AXPs in the Galactic plane.

The identification  of AXPs with magnetars was further strongly motivated
by the similarity of AXP emission to that of the soft gamma-ray
repeaters (SGRs) in quiescence.  Specifically, the latter have similar
pulse periods, are spinning down \citep{kds+98,ksh+99}, and have X-ray
spectra that are comparable to, though somewhat harder than, those of
the AXPs, at least when not in outburst \citep{mcft00,kkm+02}.
Independent evidence for the ultra-high magnetic fields exists in SGRs;
for example, a $\sim 10^{15}$~G magnetic field is required to confine
the radiation that is seen following major outbursts
\citep{td95}.  Indeed the only strong distinction between the
emission from SGRs and AXPs appeared to be the fact that SGRs exhibit
occasional and repeating short bursts, while AXPs did not.

\section*{RESULTS}

The results presented here were obtained using the Proportional Counter Array \cite[PCA;][]{jsg+96} on board NASA's {\it Rossi X-ray Timing Explorer} (\xte). The PCA consists of an array of five collimated xenon/methane multi-anode proportional counter units operating in the 2~--~60~keV range, with a total effective area of approximately $\rm{6500~cm^2}$ and a field of view of $\rm{\sim 1^o}$~FWHM.
A program to monitor AXPs regularly  using \xte\ was begun in 1996 in order to determine their long-term timing,
pulsed flux, and pulse profile stabilities \citep{kcs99,klc00,kgc+01,gk02}.  As
part of this program, motivated by the existence of SGR bursts, we also
searched the AXP data for bursts \citep{gkw02}. 
 Our observations consist primarily of short snapshots  taken on a monthly basis. In addition, we used a handful of archival observations; the exposures in these observations vary.
We used the \texttt{GoodXenonwithPropane} data mode, which records photon arrival times with 1-$\mu$s resolution and bins energies into one of 256 channels. Below, we summarize the results of this monitoring campaign thus far \cite[see][for details]{kgc+01,gk02,gkw02}.

%%%%%%%%%

\subsection*{Phase-Coherent Timing}

\begin{figure}
\begin{minipage}{0.48\textwidth}
\begin{center}
\includegraphics[width=0.95\textwidth]{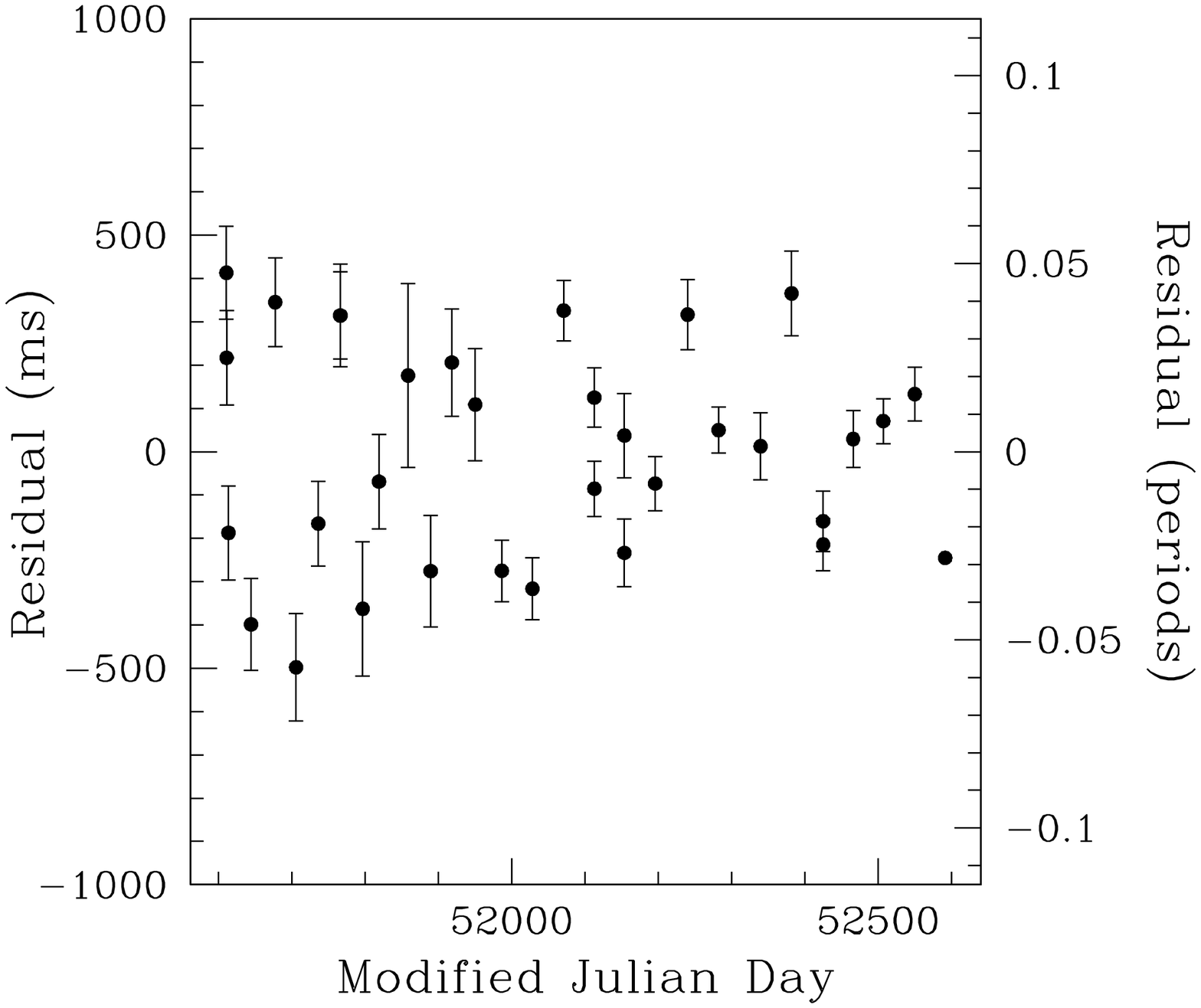}
\end{center}
\caption{Arrival time  residuals for \oft\ with $\nu$ and $\dot{\nu}$ subtracted. 
The RMS residual here is $<0.3$\% of the period, indicating
great rotational stability. However, we realize that the reduced $\chi^2$ is much greater than unity, this is under investigation. 
\label{fig:0142res}}
\end{minipage}\hfill
\begin{minipage}{0.48\textwidth}
\begin{center}
\includegraphics[width=0.85\textwidth]{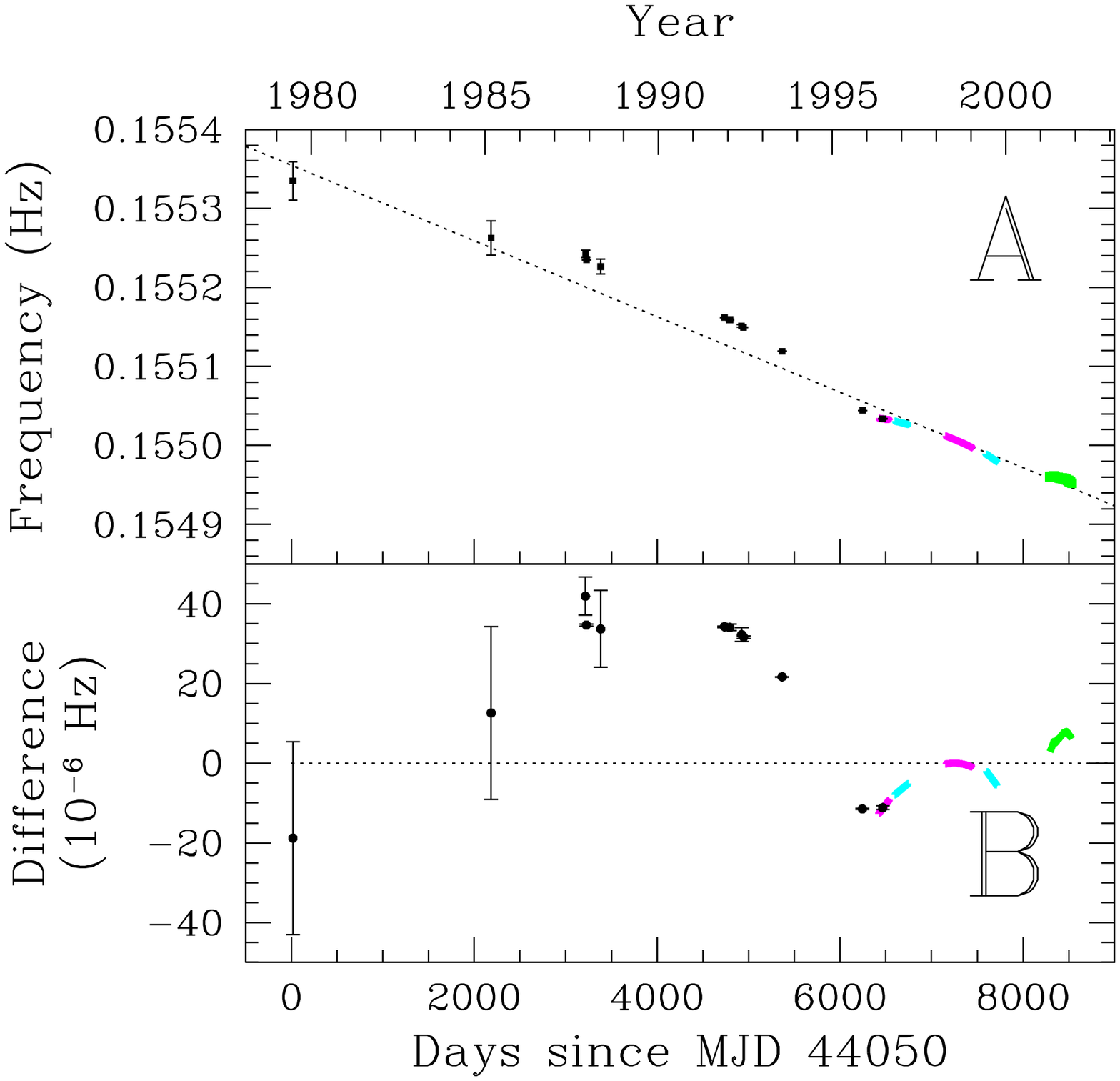}
\end{center}
\caption{Spin history for \tfe.  The points represent
past measurements of the frequency of the pulsar (see Oosterbroek et al. 1998 and references therein, Paul et al. 2000 and Baykal et al. 2000). Solid lines
represent our RXTE observations  (Kaspi et al. 2001). 
Panel A shows the observed frequencies over time.  The dotted
line is the extrapolation of a $\nu$ and $\dot{\nu}$ measured from 
a  phase-connected interval in 1999.
Panel B shows the
difference between the ephemeris indicated by the dotted line and the data points.
\label{fig:freq1048}}
\end{minipage}
\end{figure}

In the timing analysis, each binned time series was epoch-folded using the best estimate frequency determined initially from either a periodogram or Fourier transform (though later folding was done using the timing ephemeris determined by maintaining phase coherence; see below). Resulting pulse profiles were cross-correlated in the Fourier domain with a high signal-to-noise template created by adding phase-aligned profiles from previous observations.  The cross-correlation returns an average pulse time-of-arrival (TOA) for each observation corresponding to a fixed pulse phase. The pulse phase $\phi$ at any time $t$ can be expressed as a Taylor expansion,

\begin{equation}
\phi(t) = \phi(t_0) + \nu_0 (t-t_0) + \frac{1}{2} \dot{\nu}_0(t-t_0)^2   + \frac{1}{6} \ddot{\nu}_0(t-t_0)^3 + \cdots,
\label{eq:phase}
\end{equation}

\noindent where $\nu\equiv 1/P$ is the pulse frequency,  $\dot{\nu} \equiv d\nu/dt$, etc. and subscript `$0$' denotes a parameter evaluated at the reference epoch $t=t_0$. The TOAs were fit to the above polynomial using the pulsar timing software package \texttt{TEMPO}\footnote{http://pulsar.princeton.edu/tempo}. Unambiguous pulse numbering is made possible by obtaining monitoring observations spaced so that the best-fit model parameters have a small enough uncertainty to allow prediction of the phase of the next observation to within $\sim0.2$. Typically this requires two closely spaced observations  (within a few of hours of each other) followed by one spaced a few days later, and regular monitoring thereafter, as long as phase coherence can be maintained.

Using phase coherent timing we have shown that AXPs can be
phase connected over long intervals, implying they can spin down with
impressive stability,  which  is at odds with
what is seen in most X-ray pulsars that definitely have accretion disks
\cite[][see Figure~\ref{fig:0142res}]{kcs99,klc00,gk02}. 
However this is not true of all our sources.
\tfe\ shows large
deviations from a simple spin down
\cite[Figure~\ref{fig:freq1048},][]{kgc+01}.  
However,  the very noisy timing behavior seen
in \tfe\ shows no evidence for correlated flux variations, as
are expected in fall-back disk models.

\subsection*{Pulse Morphology}

Motivated by a previous report of a significant pulse morphology change in an AXP \cite[\tfn;][]{ikh92} we used our \xte\ data to search for more such events. Morphology changes are expected in the magnetar model during 
outbursts or torque variations.  

To search for pulse profile changes, the profiles were first phase aligned using the templates and the same cross-correlation procedure used for timing. Each data profile   was fit to a high signal-to-noise  template  by adjusting the amplitude and the offset in order to minimize a $\chi^2$ statistic.
The resulting data profile was subtracted from the template to yield
``profile residuals'' for each observation.
The procedure was repeated for each pulsar with different binning
 in order to have sensitivity to a variety of types
of pulse profile changes.  We have not detected any large pulse profile
variations, except for \tfn\ during a major outburst (see below). This justifies our other analysis procedures which assume a
fixed profile.  However, in a handful of observations, we have found
low-level pulse profile changes, at the $\sim 3\sigma$ level for all sources. 

It is difficult to set quantitative upper limits on the amplitude of pulse profile
changes to which we were sensitive, as these depend on the shape of the change,
and vary depending on the length of the observation.  Typically, RMS profile
residuals are $\sim$20\% of the pulse peak, although this varied from 4--40\%.

\subsection*{Pulsed Flux and Spectra}

We have also used our \xte\ observations to monitor the pulsed flux of each AXP. By considering the stability of pulsed fluxes, we test the two competing AXP models.  In accreting
systems, torque fluctuations are generally accompanied by variation in
the mass accretion rate and hence luminosity, and vice versa.  In the magnetar model, abrupt flux changes could occur in analogy with SGR-like outbursts.
Given the large field-of-view of the PCA, the low count rates for the sources relative to the background, and the fact that, for example, \tfn\ is in a supernova remnant, total flux measurements are difficult with our \xte\ data. Instead, we have determined the pulsed component of the flux, by using the off-pulse emission as a background estimator.
 
Data from each observing epoch were folded at the expected pulse period as was done for the timing analysis. However, for the flux analysis, 16 phase bins were used across the pulse. For each phase bin, we maintained a spectral resolution of 128 bins over the PCA range. Given the broad morphologies of the average pulse profile, only one phase bin could be used as a background estimator. The pulse profiles were phase aligned, so that the same off-pulse bin was used for background in every case. The remaining phase bins were summed, and their spectral bins regrouped using the \texttt{FTOOL} \texttt{grppha}. The regrouped, phase-summed data sets, along with the background measurement, were used as input to the X-ray spectral fitting software package \texttt{XSPEC}\footnote{http://xspec.gsfc.nasa.gov}. Response matrices were created using the \texttt{FTOOL}s \texttt{xtefilt} and \texttt{pcarsp}. Because of the limited statistics, fitting a two-component model was not practical, so we used a simple photoelectrically absorbed power law, at first holding only $N_H$ fixed.  For all sources we found that the photon index $\Gamma$ was constant within the uncertainties; we therefore held it fixed at its mean value.  To extract a pulsed flux at each observing epoch we refit each spectrum by varying only the normalization. Other than after a major outburst from \tfn\ (see below),  we found no evidence for any large variability in the pulsed flux within 20\%--30\% consistent with the  systematic errors. AXP spectra are generally best  fit by a two-component model consisting of a 
photoelectrically absorbed blackbody with a hard power-law tail (Israel et al. 1999). Whether 
these two components are physically distinct is an open question \citep{opk01,tlk02}. To investigate this, we
compared the pulse profile morphology of the AXPs in two energy bands. 
Thus far, we have found that pulse profile energy-dependence is generally
subtle, but varies from source to source.
We found no clear patterns when
considering average pulse profiles as a function of
energy \citep{gk02}.

%%%%%%%%%%%%%%%%%

%%%%%%%%%%%%%%%%%

%\subsection*{SGR-like X-ray Bursts from Two AXPs}

\subsection*{SGR-like X-ray Bursts from \tfe}

In the magnetar model for SGRs \citep{td96a}, bursts are a result of sudden
crustal yields due to stress from the outward diffusion of the huge internal magnetic field. Such yields cause crust shears which twist the external magnetic field, releasing energy. \citet{td96a} who, upon suggesting that AXPs are also magnetars, predicted X-ray bursts should eventually be seen from them. In order to confirm this prediction we searched our \xte\ data for bursts. Our burst searching algorithm is summarized below \cite[see ][for details]{gkw02}.

Time series  were created
with 31.25-ms resolution from photons having energies in the range
2--20~keV for each PCA Proportional Counter Unit (PCU) separately,
using all xenon layers.  Photon arrival times at each epoch were
adjusted to the solar system barycentre.  The resulting time series
were searched for significant excursions from the mean count rate by
comparing each time bin value with a windowed 7-s running mean.  Bursts
were identified assuming Poissonian statistics, and by combining
probabilities from the separate PCUs. 

\begin{figure}
\begin{minipage}{0.48\textwidth}
%\begin{figure}
\begin{center}
\includegraphics[width=0.85\textwidth]{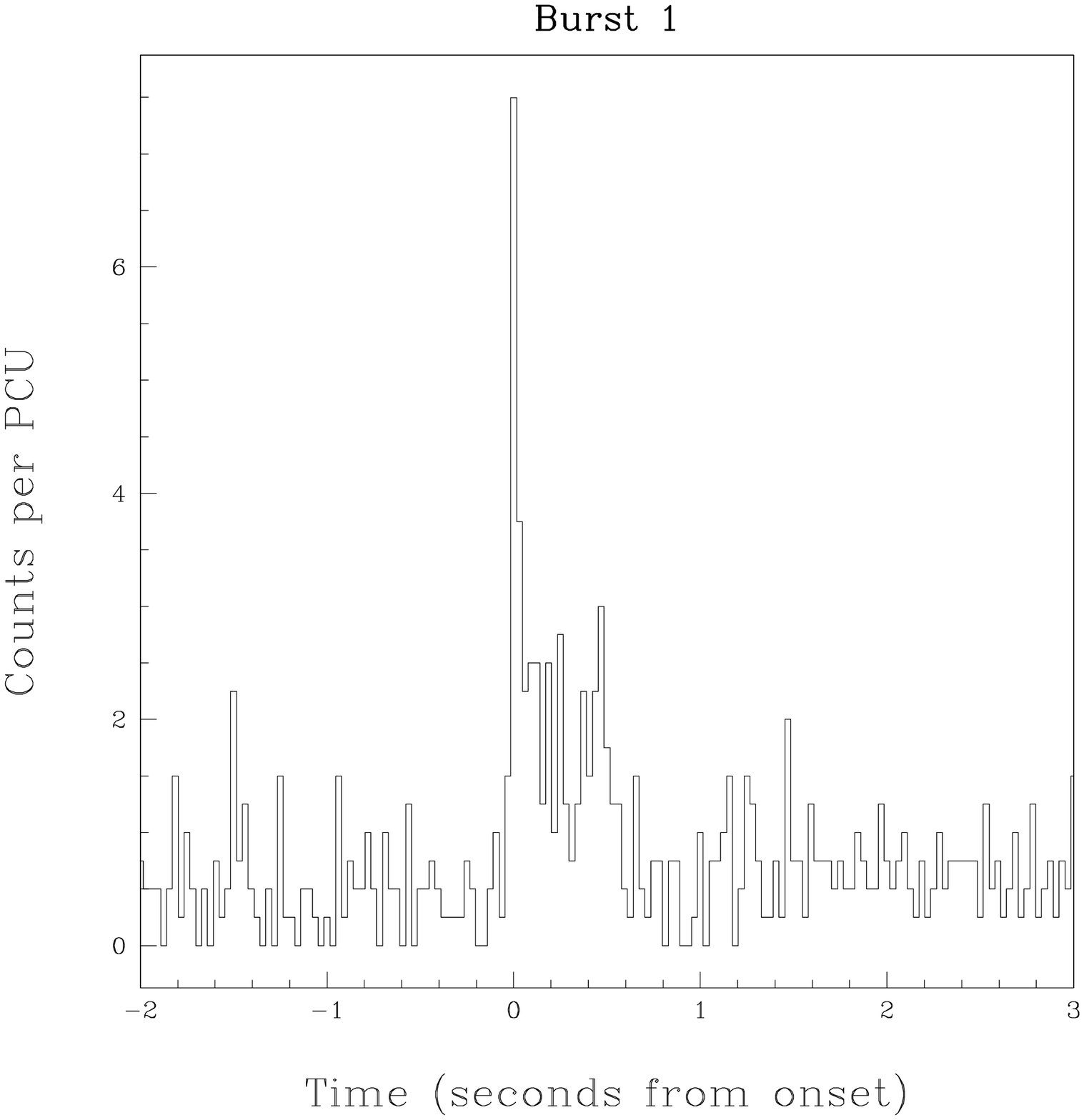}
\end{center}
%\end{figure}
\end{minipage}\hfill
\begin{minipage}{0.48\textwidth}
\begin{center}
\includegraphics[width=0.85\textwidth]{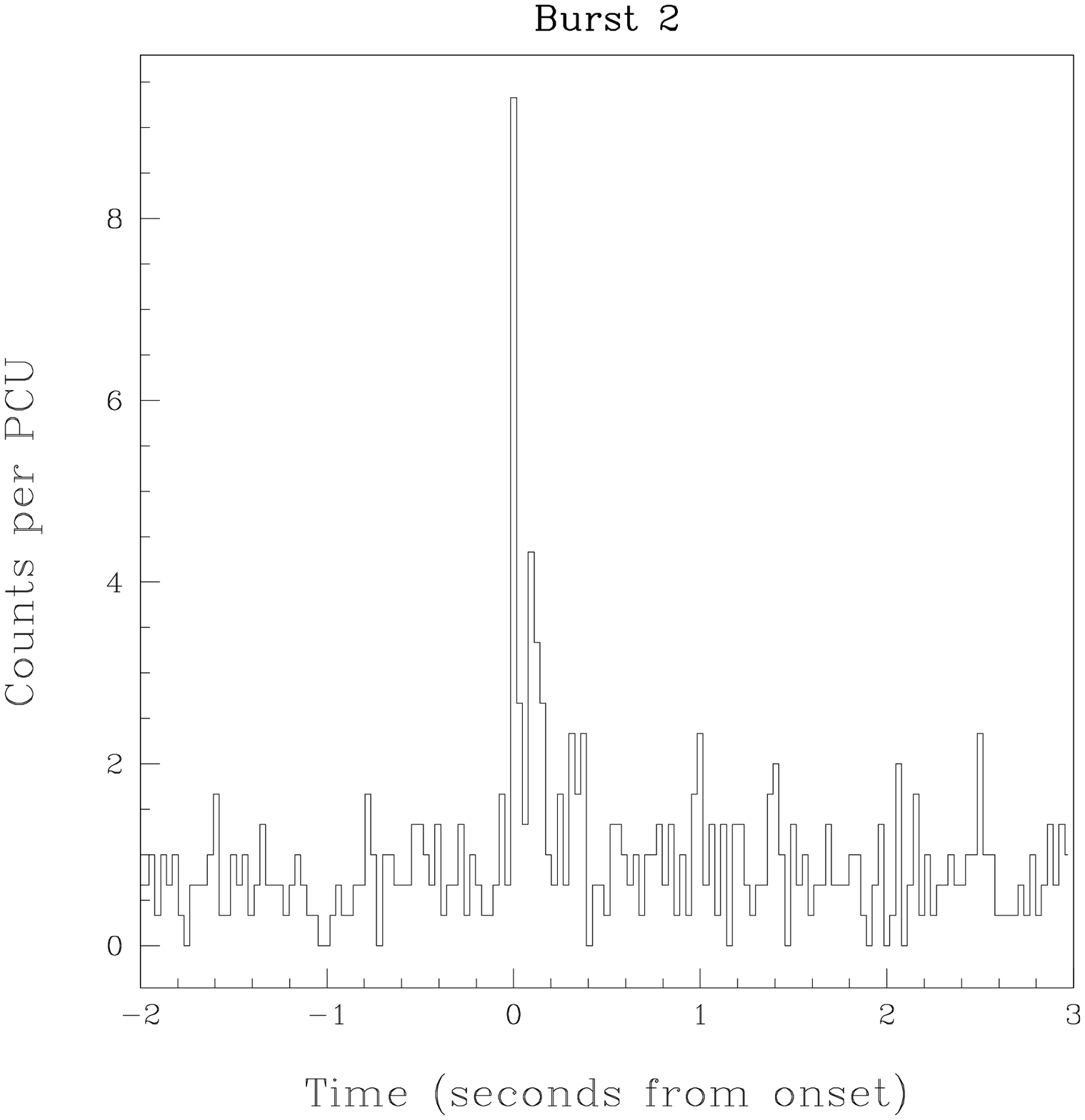}
\end{center}
\end{minipage}
\caption{
Left Panel: Background subtracted 2-20~keV light curves for  the first burst displayed with 31.25-ms time resolution. 
Right Panel: Same but for second burst. 
\label{fig:profiles}}
\end{figure}

We discovered two short, highly significant  X-ray bursts, separated by 16 days, from the direction of the AXP
\tfe\ in data from Fall 2001 \citep{gkw02}. The first occurred
during a 3-ks PCA observation obtained on 2001 October 29, and the 
second during a 3-ks
observation obtained on 2001 November 14. 
We searched a total of 442~ks of PCA observing time from 1996--2002, and no  
other significant bursts were found toward 1E~1048$-$5937.  
The burst profiles are shown in Figure~\ref{fig:profiles}.

 %  Given the PCA's 1$^{\circ}$ field-of-view, we cannot
%exclude the possibility that the bursts did not come from
%1E~1048.1$-$5937.  However no other astrophysical origin is plausible
%(Type I X-ray bursts have softer spectra, are longer, and there is no
%known LMXB in the FOV; the probability of two gamma ray bursts
%occurring within 16 days of each other in the same direction is tiny.)

%%%%%%%%%%%%%%%%%%

\begin{figure}
\begin{minipage}{0.65\textwidth}
\begin{center}
\begin{tabular}{lcc} \hline \hline
\multicolumn{1}{c}{Parameter} & \multicolumn{2}{c}{Value} \\
 &  Burst 1  & Burst 2 \\\hline
Burst day, (MJD) & 52211 &  52227\\
Burst start time, & 0.2301949(24) & 0.836323379(68) \\
 (fraction of day, UT) &&\\
Burst rise time, $t_r$ (ms) & $21^{+9}_{-5}$ & $5.9^{+2.0}_{-1.2}$ \\
Burst duration, $T_{90}$ (s) &  $51^{+28}_{-19}$ & $2.0^{+4.9}_{-0.7}$\\
$T_{90}$ fluence (counts) & 485 $\pm$ 118 & 101 $\pm$ 15 \\
$T_{90}$ fluence ($\times 10^{-10}~\mathrm{erg~cm}^{-2}$) & 20.3 $\pm$ 4.8  & 5.3 $\pm$ 1.2 \\
Peak flux for 64~ms  & $31^{+45}_{-10}$ & $26^{+23}_{-5}$ \\
($\times 10^{-10}~\mathrm{erg~s^{-1}~cm}^{-2}$) &&\\
Peak flux for $t_r$~ms   & $54^{+79}_{-17}$ & $114^{+100}_{-23}$ \\
($\times 10^{-10}~\mathrm{erg~s^{-1}~cm}^{-2}$) &&\\
\hline
power law index & $0.89^{+1.8}_{-0.71}$   & $1.38^{+0.75}_{-0.62}$ \\
power law flux & $2.0_{-1.8}^{+8.4}$ & $4.0^{+3.5}_{-0.8}$ \\
 ($\times 10^{-10}~\mathrm{erg~s^{-1}~cm^{-2}}$)   &&\\
line energy (keV) & $13.9 \pm 0.9$ & ... \\
line width, $\sigma$ (keV) & $2.2^{+1.3}_{-1.0}$ & ... \\
line flux  & $3.9_{-1.6}^{+2.2}$ & ... \\
($\times 10^{-10}~\mathrm{erg~s^{-1}~cm^{-2}}$) &&\\
reduced $\chi^2_{\nu}$ ($\nu$) & 1.24 (15) & 0.77 (5) \\
\hline\hline
\end{tabular}
\end{center}
\textsf{Table 1. AXP \tfe\ burst  properties. The burst rise $t_r$ times were determined by a maximum likelihood
fit to the unbinned data using a piecewise function having a linear
rise and exponential decay.  
The burst duration, $T_{90}$, is the
interval between when 5\% and 95\% of the total 2--20~keV burst fluence
was received.   All fluences
and fluxes are in the 2--20~keV range.  $T_{90}$ fluences in cgs units
were calculated assuming a power-law spectral model. 
For all spectral fits, $n_H$ was held fixed at $1.2\times10^{22}$~cm$^{-2}$,
the value determined from recent XMM observations \protect\citep{tgsm02}.
Spectral modeling was done using photons in the 2--40~keV
range.  
Uncertainties are 68\% confidence intervals, except for
those reported for the cgs-unit fluences and fluxes, as well as the
spectral model parameters, for which we report 90\% confidence
intervals. See \protect\citet{gkw02} for details.}
\end{minipage}\hfill
\begin{minipage}{0.32\textwidth}
\begin{center}
\includegraphics[width=0.95\textwidth]{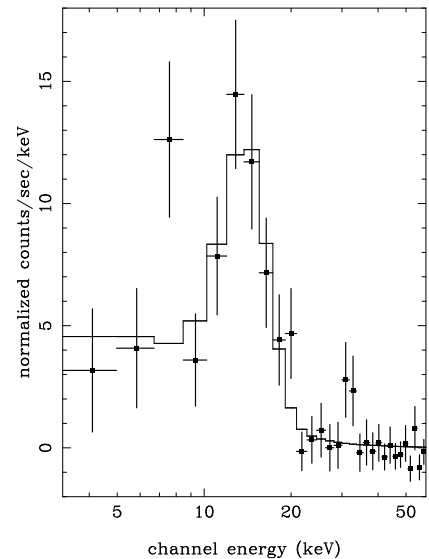}
\end{center}
\caption{
X-ray spectrum in the 2--40~keV range for the 1~s after the onset of
Burst 1. The best fit power-law plus
Gaussian line model is shown as a solid line.  The F-test shows that
the addition of a line of arbitrary energy, width and normalization to
a simple power law model improves the fit significantly, with a chance
probability of this occuring of $0.0032$.  Monte Carlo simulations in
{\tt XSPEC} were done to verify this conclusion.
\label{fig:spec1048}}

\end{minipage}
\end{figure}

We fit the spectra from the first 1~s of each burst to two
one-component models, a power law and a black body.
Continuum models provided an adequate characterization of the Burst~2
spectrum but not of the Burst~1 spectrum.  As seen in
Figure~\ref{fig:spec1048}, the spectrum for the 1~s after the Burst 1
onset exhibits a feature near 14~keV.  
This feature is clear in all binning schemes and
is prominent throughout the first $\sim$1~s of the burst.  No known PCA
instrumental effect produces a feature at this energy (K. Jahoda,
personal communication). Intriguingly, in some spectral binning schemes, we saw hints
of features in the spectrum consistent with additional multiples
of 7~keV, suggesting cyclotron lines.  However these were not
statistically significant.

Due to the wide ($\sim$1$^{\circ}$) field-of-view (FOV) and lack of
imaging capabilities of the PCA, we cannot verify that the bursts
originated from the location of the AXP.  The low peak X-ray
fluxes of the events (see Table 1) preclude determining the source's
location using data from other, better imaging instruments that were
contemporaneously observing the X-ray sky, such as the {\it RXTE} All
Sky Monitor, or the Wide Field Camera aboard {\it BeppoSAX}.
We must therefore consider other possible origins from
the bursts before concluding they were from the AXP.

The bursts' short rise times (Table~1) require emission regions of less
than a few thousand km, implying a compact object origin. 
Type I bursts from an LMXB in the same
FOV as \tfe\ are unlikely to explain our observed
bursts because (i) the burst rise times are much shorter than those of
Type I bursts; (ii) the burst spectra are much harder than those of
Type I bursts; (iii) Burst 2 shows no evidence for spectral softening
with time and no Type I burst has ever exhibited a spectral feature
like the one detected in Burst 1; (iv) the bursts are extremely faint,
implying a source location well outside the Milky Way for Type
I burst luminosities (v) there are no known LMXBs in the
FOV \citep{lph01}.  
Furthermore,  the bursts we have observed are unlikely to
be Type II bursts from an unknown X-ray binary in the PCA FOV because (i)
of the rarity of such events (observed thus far only in two sources, both accreting binaries); 
(ii) Type II bursts have longer rise
times than do our bursts; (iii) no Type II burst has exhibited a
spectral feature like that seen in Burst 1.
Classical gamma-ray bursts (GRBs) sometimes exhibit prompt X-ray
emission that can have temporal and spectral signatures similar to
those we have observed\citep{hzkw01}.  However, the likelihood of two
GRBs occurring within $1^{\circ}$ of each other is small, and GRBs are
not known to repeat.

The observed burst properties are in many ways similar to those seen
from SGRs \citep{gkw+01}.  The fast rise and slow decay profiles are
consistent with SGR time histories, as are the burst durations.   
Both AXP and SGR
bursts are spectrally much harder than is their quiescent pulsed
emission.  The burst peak fluxes and fluences fall within the range
seen for SGRs, and the spectrum of Burst 2 is consistent with SGR burst
spectra of comparable fluence.  Burst 1 has characteristics unlike
nearly all SGR bursts, specifically its long tail (Table~1) and spectral
feature.  However, we note that one event from SGR~1900+14 was
shown \citep{isw+01,si00} to possess each of these properties.  The
marginal increase in the pulsed flux that we observed at the burst
epochs is consistent with SGR pulsed flux increases seen during bursting
episodes \citep{wkg+01}.  Finally, the fact that in spite of several years
of monitoring, the only two bursts detected occurred within two weeks of
each other suggests episodic bursting activity, the hallmark of SGRs. 
Thus, the characteristics of these events match the burst properties of
SGRs far better than any other known burst phenomenon.

The large 14-keV spectral line in the first burst in intriguing. 
An electron cyclotron feature at this energy
$E$ implies a magnetic field of $B = 2\pi m c E / h e \simeq 1.2 \times
10^{12}$~G (where $m$ is the electron mass, $c$ is the speed of light,
$h$ is Planck's constant, and $e$ is the electron charge), while a
proton cyclotron feature implies $B \simeq 2.4 \times 10^{15}$~G.  The
former is significantly lower than is implied from the source's
spin-down and is typical of conventional young neutron stars, rather
than magnetars.  The latter is higher than is implied by the spin-down
yet reasonable for the magnetar model as the spin-down torque is
sensitive only to the dipolar component of the magnetic field.

\subsection*{A Major Outburst from \tfn}

%\subsubsection*{A Major Outburst from \tfn}

\begin{figure}
\begin{minipage}{0.48\textwidth}
\begin{center}
\includegraphics*[width=0.85\textwidth,height=3in]{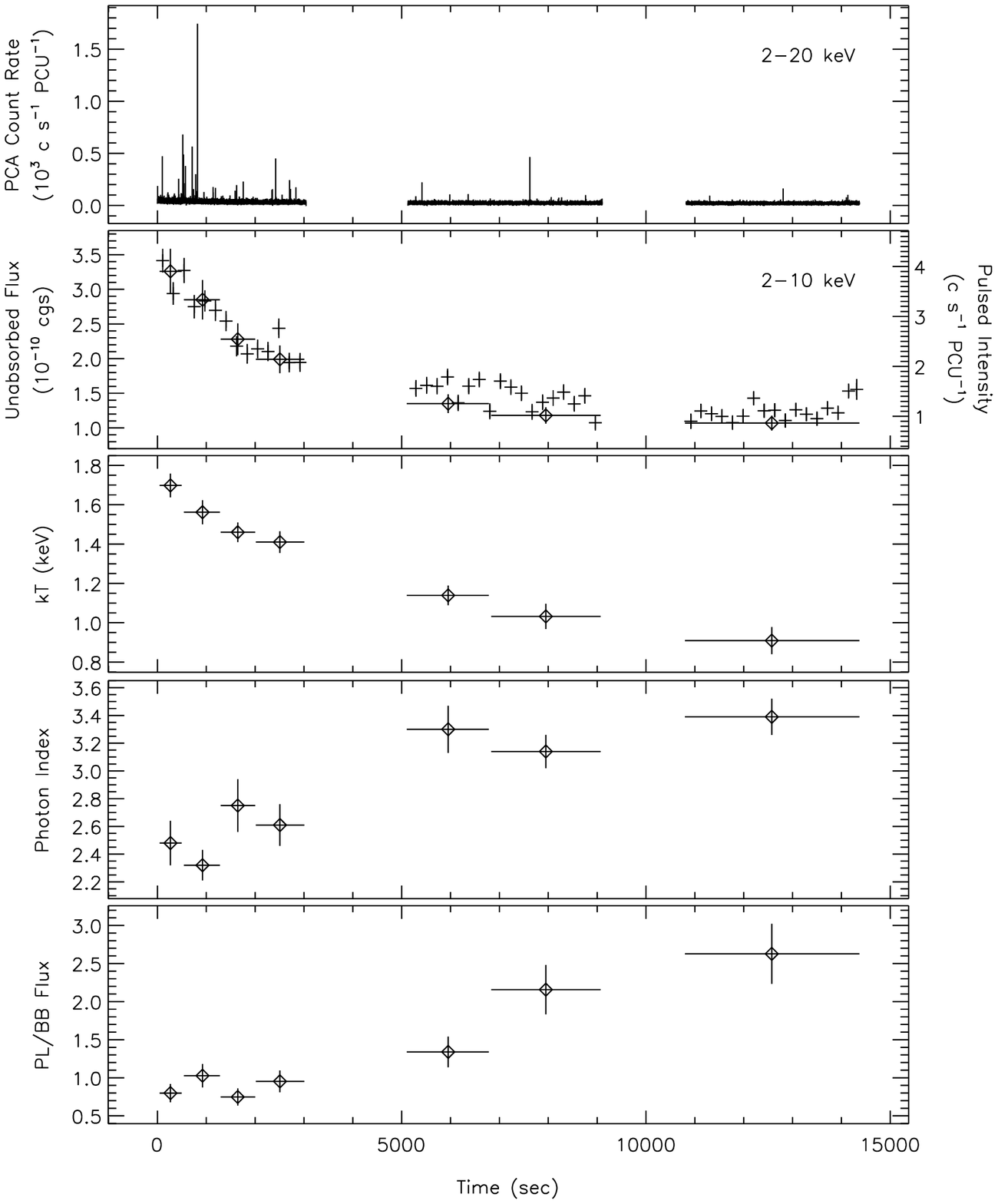}
\end{center}
\caption{
The time evolution of 1E 2259+586 during its burst active
episode on 2002 June 18 as observed with the RXTE PCA.  Top: 
lightcurve in 2-20 keV X-rays with 0.125~s time
resolution. Bottom: The pulsed 2--10~keV flux. The crosses represent the pulsed flux in the 2--20~keV band and the diamonds represent the unabsorbed persistent flux.
\label{fig:b2259}}
\end{minipage}\hfill
\begin{minipage}{0.48\textwidth}
\begin{center}
\includegraphics[width=0.65\textwidth]{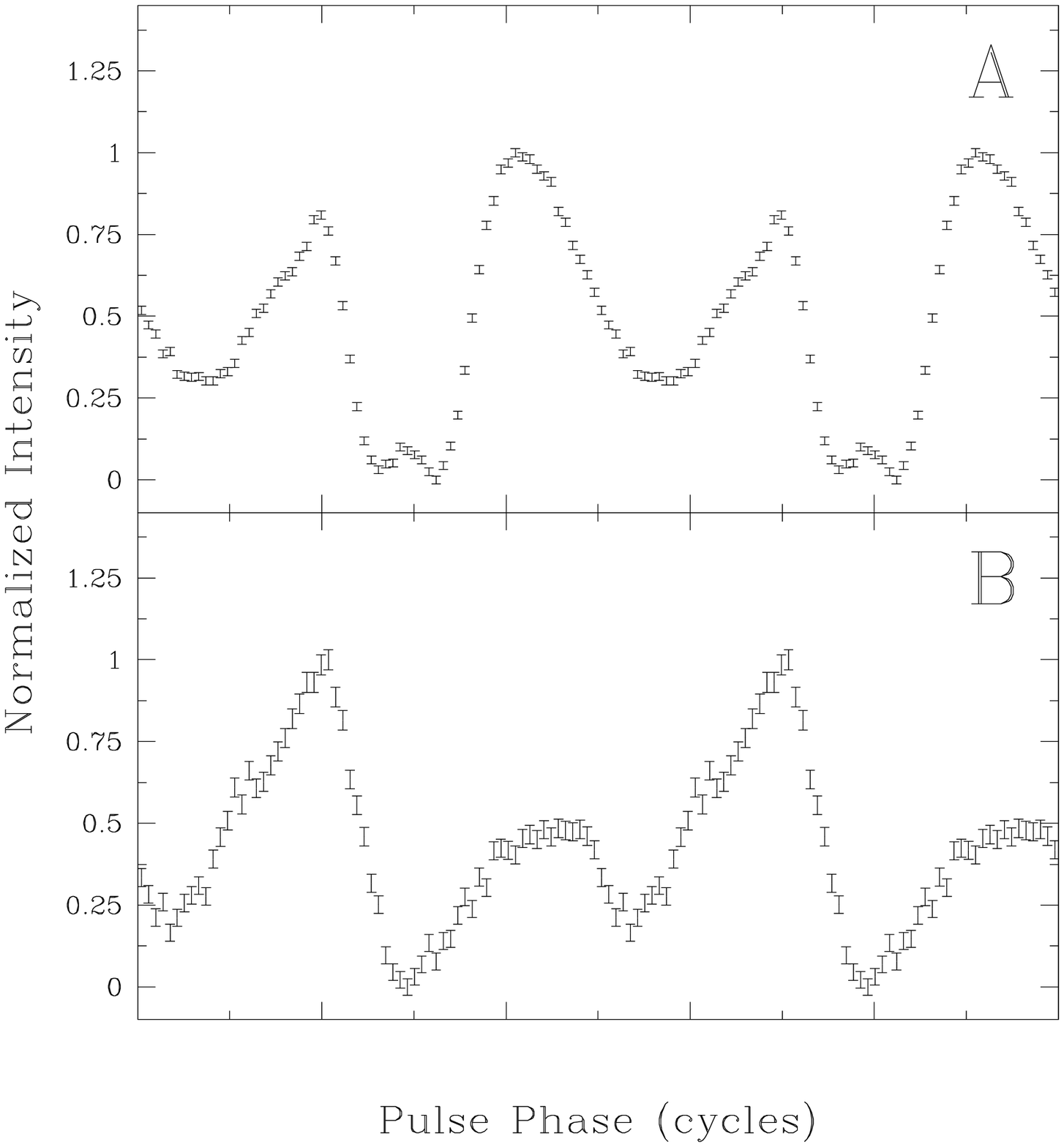}
\end{center}
\caption{
Average pulse profiles of \tfn\ in the 2.5--9.0~keV band. Two cycles are plotted for clarity. 
(A) Average profile before the outburst (total exposure time: 764~ks). (B) Average profile during the outburst, with bursts omitted (total exposure time 11~ks). During the outburst, the amplitudes of the peaks relative to the pre-outburst profile are clearly reversed. The relative phase displayed is that successfully used in our timing analysis. 
\label{fig:pulse2259}}
\end{minipage}
\end{figure}

On 2002 June 18, fortuitously during one of our {\it RXTE} monitoring
observations  of \tfn, 
over 80 bursts were detected in the span of 14~ks \citep{kg02,kgw02}.  
Figure~\ref{fig:b2259} shows the 0.125~s light curve.
\tfn, in contrast to \tfe, had been an extremely stable
rotator and had shown no signs of ``volatility.''
The association of the bursts toward \tfe\ with the AXP could not be 
unambiguously verified, however,  the bursts seen here  (Figure~\ref{fig:b2259}; top panel) clearly originated from \tfn. 
During and following the outburst, the pulsed flux  increased 
(see Figure~\ref{fig:b2259})  and there was a significant pulsed morphology change, such that the relative amplitude of the leading and 
trailing pulse were swapped (see Figure~\ref{fig:pulse2259}).

Target-of-opportunity observations were made by many major observatories
in addition to {\it RXTE}
at the time of this outburst.  No radio emission was detected in spite
of sensitive {\it VLA} observations, nor were any bursts seen in TOO {\it XMM-Newton}
data obtained roughly a week afterward.  Near-infrared observations
obtained with the Gemini North 8-m reflector on June 21 showed that the
proposed IR counterpart had brightened by a factor of 3.4
(Kaspi et al. 2002).  A few days later, it faded back to its pre-burst 
value.  This represents the first detection of transient optical/IR emission 
from an isolated neutron star.  Subsequently, \citet{ics+02} 
reported possible IR variability from the proposed
counterpart to 1E~1048.1$-$5937, which they suggested might be related
to its bursting behaviour.  The origin of IR emission is currently not
addressed in the magnetar model.

%\subsubsection*{Pulse Morphology and Pulsed Flux}

%%%%%%%%%%%%%%%%%%%%%%%%%%%%%%%%%%%%%%%%%%%%%%%%%%%%%%%%%%%%%%%%%%%%%%%%%%%
%\begin{figure}[t]
%\centerline{\psfig{file=flux1841.ps,width=0.4\textwidth}}
%\caption{Pulsed flux time series for \xte\ observations of \efo. Error bars represent 1$\sigma$ confidence intervals.
%\label{fig:flux1841}}
%\end{figure}
%%%%%%%%%%%%%%%%%%%%%%%%%%%%%%%%%%%%%%%%%%%%%%%%%%%%%%%%%%%%%%%%%%%%%%%%%%%

Pulse morphology changes can occur, according to the magnetar model, during
 outbursts as in the SGRs, as this is when 
magnetic reconfigurations occur.  Indeed during the outburst of \tfn, 
significant pulse morphology changes were seen, such that the relative
amplitude of the double peaked profile was swapped (see Figure~\ref{fig:pulse2259}). 
\citet{ikh92} also
reported a significant change in the pulse morphology of \tfn\ in
\textit{GINGA} observations obtained in 1990, such that the leading
pulse had amplitude roughly half of the trailing pulse.  In both
instances the pulse profile returned
back to its long-term state.

Pulsed flux enhancements in both the pulsed and persistent flux 
are observed in the SGRs during bursting episodes and are
thought to be a result of back heating of the surface and outer crust 
by the burst emission in the magnetosphere.  The
pulsed flux of \tfn\ increased by a factor of $\sim 4$ during its bursting
episode (see Figure~\ref{fig:b2259}).  
We noted a marginal ($\sim 3\sigma$) increase
in the pulsed flux from 1E~1048.1$-$5937 that commenced with the
observation in which Burst~1 was detected, and which lasted
$\sim$4~weeks \citep{gkw02}.  There have been past claims of X-ray flux
changes, by up to a factor of $\sim$10 in two AXPs \cite[\tfn\ and
\tfe;][]{bs96,opmi98}, although the reported fluxes were measured with
different instruments having very different spectral responses and
angular resolutions. If correct, these past claims of increased flux
might imply previous bursting activity.

\section*{DISCUSSION}

We have presented a variety of observational results for all AXPs 
using regular monitoring observations and archival observations from 
the PCA aboard {\it RXTE}. There is now strong evidence that AXPs are magnetars. 
The great rotationally stability in some AXPs is at odds with an accretion scenario, as is  
the absence of pulsed flux or pulse morphology changes correlated with torque. 
Extended periods of spin-up would only, though not necessarily,  be expected in an accretion scenario, however all AXPs 
have exhibited  relatively steady spin-down. Furthermore, and perhaps most compellingly, accretion models cannot account for the bursting episodes seen in two AXPs. Interestingly, there is a correlation between timing stability
and $\dot{\nu}$ in our data.  The sources with by far the smallest
$\dot{\nu}$'s, are the most stable in quiescence (at
least during our observations), while those with large $\dot{\nu}$'s 
are less so. The SGRs, being even less stable \citep{wkg+02},
have even larger $\dot{\nu}$'s, in agreement with this trend.  
If correct, the  trend suggests a continuum of timing
properties between the AXP and SGR populations, lending additional
support to the connection between them. 

The identification with magnetars is further strongly motivated by the
similarity of the AXP emission to that of the SGRs. 
Specifically, the latter have similar pulse
periods, are spinning down \citep{kds+98,ksh+99}, and have X-ray
spectra that are comparable to, though somewhat harder than, those of
the AXPs, at least when not in outburst. 
With the detection of X-ray bursts from two AXPs the two classes
are now similar  in that respect as well.
These very recent results appear to have sealed the connection between SGRs and AXPs, as was originally predicted by \citet{td96a}. However, the identification of the two classes objects as magnetars has left some open questions. For instance, what if anything distinguishes AXPs and SGRs? Are AXPs the progenitors of SGRs or vice versa? Recently \citet{kkm+02} have shown that the X-ray spectrum of the quiescent counterpart to SGR~0526--66 has a power-law index $-3$, similar to those seen in AXPs, but softer than those of SGRs. As SGR~0526--66 has been inactive for almost two decades, this suggests SGRs and AXPs are one and the same object, and may have even cycle back and forth between each flavor.
Another noteworthy question is what is the magnetar/radio pulsar connection? Recent observations of radio pulsars having apparent surface dipolar magnetic fields comparable \citep{cml+00}, and in one very new case (McLaughlin et al., private communication), surpassing  those of some AXPs, leaves open the question of what physically distinguishes a magnetar from a rotation-powered pulsar of high magnetic field? \citet{pvc00} suggested that  the sources classified as magnetars might have much higher fields, in the form of multipole moments.
Perhaps comparing the thermal X-ray emission from radio pulsars having a variety of magnetic fields may  illuminate the situation.

\section*{ACKNOWLEDGMENTS}

This work was supported by
NASA, NATEQ and CIAR. 
VMK is a Canada Research Chair and a Fellow of the CIAR.  This
research has made use of data obtained through the High Energy
Astrophysics Science Archive Research Center Online Service, provided
by the NASA/Goddard Space Flight Center.

\bibliographystyle{apalike}
%\bibliography{journals1,modrefs,psrrefs,crossrefs}

\vspace*{2\baselineskip}
\noindent E-mail address of V.M. Kaspi \ \ \  \url{vkaspi@physics.mcgill.ca}

\end{document}